\documentclass[aps, prb, twocolumn, showpacs]{revtex4}
\usepackage{amsmath, amsfonts, amssymb}
\usepackage{bm}
\usepackage{hyperref}
\usepackage{graphicx}

\begin{document}

\def\Re {\mbox{Re}}
\def\Im {\mbox{Im}}
\newcommand{\avg}[1]{\langle#1\rangle}
\newcommand{\odiff}[2]{\frac{\di #1}{\di #2}}
\newcommand{\pdiff}[2]{\frac{\partial #1}{\partial #2}}
\newcommand{\di}{\mathrm{d}}
\newcommand{\ii}{i}
\newcommand{\norm}[1]{\left\| #1 \right\|}
\renewcommand{\vec}[1]{\mathbf{#1}}
\newcommand{\ket}[1]{|#1\rangle}
\newcommand{\bra}[1]{\langle#1|}
\newcommand{\pd}[2]{\langle#1|#2\rangle}
\newcommand{\tpd}[3]{\langle#1|#2|#3\rangle}
\renewcommand{\vr}{{\vec{r}}}
\newcommand{\vk}{{\mathbf{k}}}
\renewcommand{\ol}[1]{\overline{#1}}

\title{Stable topological superconductivity in a family of two-dimensional  fermion models}
\author{Meng Cheng}
\affiliation{Condensed Matter Theory Center and Joint Quantum Institute,
Department of Physics, University of Maryland, College Park, MD 20742}
\author{Kai Sun}
\affiliation{Condensed Matter Theory Center and Joint Quantum Institute,
Department of Physics, University of Maryland, College Park, MD 20742}
\author{Victor Galitski}
\affiliation{Condensed Matter Theory Center and Joint Quantum Institute,
Department of Physics, University of Maryland, College Park, MD 20742}
\author{S. Das Sarma}
\affiliation{Condensed Matter Theory Center and Joint Quantum Institute,
Department of Physics, University of Maryland, College Park, MD 20742}

\date{\today}

\begin{abstract}
We show that a large class of two-dimensional spinless fermion models exhibit topological superconducting phases characterized by a non-zero Chern number. More specifically, we consider a generic one-band Hamiltonian of spinless fermions that is invariant under both time-reversal, $\mathbb{T}$, and a group of rotations and reflections, $\mathbb{G}$, which is either the dihedral point-symmetry group of an underlying lattice, $\mathbb{G}=D_n$,
or the orthogonal group of rotations in continuum, $\mathbb{G}={\rm O}(2)$.
Pairing symmetries are classified according to the irreducible representations
of $ \mathbb{T} \otimes \mathbb{G}$. We prove a theorem that for any
two-dimensional  representation of this group, a time-reversal symmetry breaking paired state is energetically favorable. This implies that the ground state of any spinless fermion Hamiltonian in continuum or on a square lattice with a singly-connected Fermi surface
 is always a topological superconductor in the presence of attraction in at least one channel.
Motivated by this discovery, we examine phase diagrams of two specific lattice models with nearest-neighbor hopping and attraction on a square lattice and a triangular lattice. In accordance with the general theorem, the former model exhibits only a topological $(p + ip)$-wave state, while the latter shows a doping-tuned quantum phase transition from such state to a non-topological, but still exotic $f$-wave superconductor.
\end{abstract}
\pacs{74.20.Rp, 05.30.Pr, 03.67.Pp, 71.10.Fd}
\maketitle

\section{Introduction}
\label{sec:intro}
The field of quantum condensed matter has provided us with quantum many-body states that are nothing short
of amazing. Among the most remarkable are phases  associated with a new paradigm~\cite{nayak_RevModPhys'08} of
topological order. These topological phases have a number of fascinating and technologically useful
properties, such as quantized Hall conductance and non-Abelian quasi-particles.
However, the precise conditions for a quantum topological phase to arise from a microscopic Hamiltonian are unknown.
The few known examples of topological order currently include the quantum Hall states~\cite{Thouless1982},
lattice spin models due to Kitaev~\cite{Kitaev2006}, lattice versions of the quantized Hall
effect~\cite{Haldane1988}, and related models of topological insulators~\cite{Kane2005}.
While the physical systems that may host the nontrivial topological phases are quite different,
their theoretical descriptions often formally reduce to that of a topological superconductor (SC).
{\em E.~g.}, the Moore-Read even-denominator fractional quantum Hall wave-function is equivalent to
the mean-field BCS state of a spinless $(p + ip)$-wave SC \cite{read_prb'00}.
Topological insulators and SCs too can be related and classified on an equal
footing~\cite{Schnyder2008,Kitaev2009}, by noticing that any SC is an insulator of its
Bogoliubov excitations, whose ``band structure'' is governed by Bogoliubov-de
Gennes Hamiltonian at the mean-field level. Therefore, understanding topological
superconductivity is an important issue both due to its many connections to a variety
of seemingly unrelated topological phases and also in its own right, {\em e.~g.}, in
relation to the recent experimental observation of an exotic paired state in
Sr$_2$RuO$_4$~\cite{Xia2006,Kidwingira2006} and proposals for realization of
$p$-wave superfluids in cold atom systems~\cite{Gurarie20072}.

Topological SCs, most notably $(p + ip)$ models, have been
considered in the theoretical literature in great detail. However, the starting point of all theoretical models has  been a quadratic mean-field Hamiltonian, with a predetermined topological order parameter of interest, or equivalently a reduced BCS Hamiltonian with exotic interactions that are difficult to imagine being realized in the laboratory. Such models are capable of answering some key questions related to the properties of a given topological phase, but they do
not provide much guidance in the search of Hamiltonians that
would host those phases. {In other words, these models are sufficient
to produce nontrivial topological order by design, but
do not shed light on the minimal necessary conditions for the
emergence of topological order.}

In this paper we  prove a general theorem that allows us to construct a large family of lattice models that give rise to topological superconducting states. We show that contrary to a common perception, the nontrivial topological phases do not necessarily arise from exotic Hamiltonians, but instead appear naturally within a range of simple models of spinless (or spin-polarized) fermions with physically reasonable interactions. Our theorem is based on examining the BCS free energy of possible paired states {which is known to be asymptotically exact for weak coupling since BCS instability is an infinitesimal instability} and the use of the
Jensen's inequality, which ensures that topological phases are often selected
naturally by energetics. The paper is organized as follows: In Section~\ref{sec:model}, we introduce a general microscopic Hamiltonian describing spinless fermions in two dimensional space and present the BCS mean field treatment of superconductivity in this model. The topological classification of 2D superconductors is reviewed in Section \ref{Class} and the main energetics argument indicating that the topological paired states are energetically favorable is proven in Section~\ref{Theorem}. In Section~\ref{sec:lattice}, we  quantitatively study phase diagrams of two specific lattice fermion Hamiltonians on a square lattice and triangular lattice with nearest-neighbor hoppings and interactions. The ground state of the square-lattice model is proven to be a topological $(p + ip)$-wave SC at arbitrary filling. The triangular lattice model gives rise to a $(p + ip)$-wave superconducting state guaranteed at low filling, but shows a first-order  phase transition into a non-topological $f$-wave SC at intermediate fillings and another transition within the $f$-wave superconducting dome from a gapless to a fully gaped superconductor.

\section{Spinless Fermion Superconductivity in Two dimensions}
\label{sec:model}
We start our general discussion with the following single-band Hamiltonian for spinless fermions
\begin{align}
\label{H0}
\hat {\cal H}\!=\!\!\int\limits_{\mathbf{k} \in {\rm BZ}} \xi_\mathbf{k}\hat{c}^\dagger_\mathbf{k} \hat{c}_\mathbf{k}
+ \frac{1}{2}\!\int\limits_{{\mathbf{q}/2},\mathbf{k},\mathbf{k}' \in {\rm BZ}}  f_{\mathbf{k}\mathbf{k}',\mathbf{q}}
\hat{c}^\dagger_{\mathbf{k}+\mathbf{q}} \hat{c}^\dagger_{-\mathbf{k}}
\hat{c}_{-\mathbf{k}'} \hat{c}_{\mathbf{k}'+\mathbf{q}},
\end{align}
where $\hat{c}^\dagger_{\mathbf{k}}$/ $\hat{c}_{\mathbf{k}}$ are the fermion creation/annihilation
operators corresponding to momentum $\mathbf{k}$, ``${\rm BZ}$'' stands for ``Brillouin zone,''
$\xi_\mathbf{k}=\epsilon_\mathbf{k}-\mu$ with $\epsilon_\mathbf{k}$ being the dispersion relation of the
fermions and $\mu$  the chemical potential, and $f_{\mathbf{k},\mathbf{k}',\mathbf{q}}$
describes an interaction, which is assumed to have an attractive channel.

We assume that Hamiltonian (\ref{H0}) arises from a real-space lattice or continuum model and is invariant with respect to the underlying spatial symmetry group, which we denote as $\mathbb{G}$, and the time-reversal group,  $\mathbb{T}$. We note that in two dimensions (2D) the range of possible
spatial groups,  $\mathbb{G}$,  is limited to the following dihedral point-symmetry groups: $D_1$, $D_2$, $D_3$, $D_4$, and $D_6$ in the case of a lattice or orthogonal group of rotations ${\rm O}(2)=D_\infty$
in continuum. We recall that the group $D_n$ includes  $\frac{360\,^{\circ}}{n}$-rotations and in-plane reflections with respect to $n$ axes. The superconducting order parameter is classified
according to the irreducible representations of the full group $\mathbb{T} \otimes \mathbb{G}$. Since, $\mathbb{T} = \mathbb{Z}_2$, $\mathbb{Z}_2 \otimes D_1 = D_2$ and $\mathbb{Z}_2 \otimes D_3 = D_6$, we can confine ourselves to studying representations of $D_2$, $D_4$, $D_6$, and ${\rm O}(2)$, which exhaust all physically relevant possibilities.

\subsection{BCS Mean Field Theory}
\label{BCSMF}

Now, we define the superconducting order parameter as
\begin{equation}
\Delta_\mathbf{k}=\int\limits_{\mathbf{k'} \in {\rm BZ}} \tilde{f}_{\mathbf{k},\mathbf{k}'}\avg{\hat{c}_\mathbf{-k'}  \hat{c}_\mathbf{k'}},
\label{def_op}
\end{equation}
where $\tilde{f}_{\mathbf{k},\mathbf{k}'}=(f_{\mathbf{k},\mathbf{k}',\mathbf{q}=0}-f_{\mathbf{k},-\mathbf{k}',\mathbf{q}=0})/2$
is the antisymmetrized BCS coupling strength.
Using a Hubbard-Stratonovich decoupling in Eq.~(\ref{H0})
with ${\bf q} = {\bf 0}$ and ignoring superconducting fluctuations, we arrive at
\begin{equation}
\begin{split}
\hat{\cal H}^{\textrm{MF}} = \int\limits_{\mathbf{k} \in {\rm BZ}} \Big(\xi_\mathbf{k}\hat{c}^\dagger_\mathbf{k} \hat{c}_\mathbf{k}
+&
\frac{1}{2}\Delta_{\mathbf{k}}\hat{c}^\dagger_\mathbf{k} \hat{c}^\dagger_\mathbf{-k} +  \frac{1}{2}\Delta^*_{\mathbf{k}}\hat{c}_\mathbf{-k} \hat{c}_\mathbf{k}\Big)  \\
&-\frac{1}{2}\int\limits_{\mathbf{k},\mathbf{k}' \in {\rm BZ}}
\Delta_{\mathbf{k}}^* \tilde{f}^{-1}_{\mathbf{k},\mathbf{k}'} \Delta_{\mathbf{k}'},
\end{split}
\label{HBCS}
\end{equation}
with $\tilde{f}^{-1}_{\mathbf{k},\mathbf{k}'}$ being the matrix inverse of $\tilde{f}_{\mathbf{k},\mathbf{k}'}$.
By integrating out the fermions we find the
BCS free energy functional expressed in terms of $\Delta$. It contains two parts,
${\cal F}\left[\Delta_{\mathbf{k}}\right]= {\cal F}_{I} + {\cal F}_{II}$, with
\begin{align}
\label{F1}
&{\mathcal F}_I \left[\Delta_{\mathbf{k}}\right] = \!-\!T\int_{\mathbf{k} \in {\rm BZ}} \ln\! \left[2\cosh\frac{1}{2T}\sqrt{\xi_\mathbf{k}^2+ |\Delta_\mathbf{k}|^2}\right] \\
\label{F2}
&{\mathcal F}_{II}\left[\Delta_{\mathbf{k}}\right] =-\frac{1}{2}
\int_{\mathbf{k},\mathbf{k}' \in {\rm BZ}}
\Delta_{\mathbf{k}}^* \tilde{f}^{-1}_{\mathbf{k},\mathbf{k}'} \Delta_{\mathbf{k}'}.
\end{align}

\subsection{Topological Classification of Two-dimensional Superconductors}
\label{Class}

To describe the topological properties of a SC state, we introduce
the  topological index (the Chern number) as follows~\cite{volovik_JETP'88}
\begin{align}
C=\int\limits_{\mathbf{k} \in {\rm  BZ}}\frac{d^2\mathbf{k}}{4\pi}\,\vec{m}\cdot \partial_{k_x}\vec{m}\times \partial_{k_y}\vec{m},
\label{eq:chern_number}
\end{align}
where  $\vec{m} \equiv (m_1, m_2, m_3) = (\Re \Delta_{\mathbf{k}}, - \Im \Delta_{\mathbf{k}},\xi_\mathbf{k})/E_{\mathbf{k}}$ and $E_{\mathbf{k}}=\sqrt{\xi_{\mathbf{k}}^2+|\Delta_{\mathbf{k}}|^2}$. This topological index classifies all maps from $T^2$ to $S^2$ representing the unit vector $\vec{m}(\vec{k})$ into equivalent homotopy classes. We will call a SC state topological, if $C \ne 0$.

The Chern number is equal to the sum of the winding numbers, $C = \sum_\sigma W_\sigma$, which can be defined for each segment of the Fermi surface (FS), ${\cal P}_\sigma$, as follows:
\begin{equation}
2 \pi W_\sigma=\oint_{\mathcal{P}_\sigma} \nabla_\mathbf{k} \varphi_{\mathbf{k}}\cdot d \mathbf{k}
\end{equation}
where $\varphi_{\mathbf{k}}$ is the complex phase of $\Delta_\mathbf{k}$. Note that even though we assume a single-band picture, a general situation is allowed where the FS is formed by one or more disconnected components, ${\rm FS}\,=\sum_\sigma\mathcal{P}_{\sigma}$ with $\sigma=1,2.\ldots,n$.

To prove the relation between $C$ and $W_\sigma$'s, we separate the closed Brillouin zone, $\partial ({\rm BZ}) = \partial \left(S^1 \times S^1\right) = 0$, into an ``electron'' region, $E_{\rm BZ} = \left\{ {\bf k} \in {\rm BZ}:\, m_3({\bf k})>0 \right\}$ and a ``hole'' region,  $H_{\rm BZ} = \left\{ {\bf k} \in {\rm BZ}:\, m_3({\bf k})<0 \right\}$.  The Fermi surface is a directed  boundary of these regions,  ${\rm FS} = \sum_\sigma \mathcal{P}_\sigma = {\partial} E_{\rm BZ} = - {\partial} H_{\rm BZ}$. One can show that
\begin{align}
C= \frac{1}{2} \left[\,\, \int\limits_{{\bf k} \in E_{\rm BZ}} - \int\limits_{{\bf k} \in H_{\rm BZ}} \right]  \boldsymbol{\nabla}_\mathbf{k}
\!\times\! \left[\frac{m_1 \boldsymbol{\nabla}_\mathbf{k} m_2
\!-\!m_2 \boldsymbol{\nabla}_\mathbf{k}m_1}{1+|m_3|}\right].
\label{CW}
\end{align}
Eq.~(\ref{CW}) and the Stoke's theorem~\cite{Haldane2004} yield $C = \sum_\sigma W_\sigma$.

If $W_\sigma=0$ for all $\sigma$, the complex phase of the pairing order parameter
can be gauged away via a non-singular redefinition of the fermion fields and corresponds to a topologically trivial state. This however is impossible
if at least one winding number is non-zero. We will call such states time-reversal-symmetry breaking
(TRSB) states. The class of TRSB superconductors is larger than and includes that of closely related topological SCs.
If there is just one singly-connected FS, the two types of states are equivalent.

\subsection{General Theorem of the Stability of TRSB SC States}
\label{Theorem}

Now we examine the stability of TRSB SCs. The order parameter in a certain  channel corresponding to a $d_{\Gamma}$-dimensional irreducible representation, $\Gamma$, of the group $\mathbb{T} \otimes \mathbb{G}$ can be written as a linear combination of {\em real} eigenfunctions of $\Gamma$, $\phi_{a}^{\Gamma}({\bf k})$ (with $a = 1, \ldots, d_{\Gamma}$)
\begin{equation}
\label{Delta}
\Delta_{\mathbf{k}} =\sum\limits_{a=1}^{d_{\Gamma}} \lambda_a \phi_{a}^{\Gamma}({\bf k}).
\end{equation}
In two dimensions, the number of irreducible representations  to be considered is highly constrained and includes only 1D and 2D real representations. In particular: (i)~For a system with a four-fold rotational symmetry ({\em e.~g}., arising from a square lattice), the corresponding point group, $D_4$, has only one space-inversion-odd irreducible representation, $E$, which is two-dimensional; (ii)~With a six-fold rotational symmetry ({\em e.~g.}, due to a triangular or hexagonal lattice), there exist three irreducible representations of $D_6$ odd under space inversion: A 2D representation, $E_1$ (corresponding to a $p$-wave pairing) and two 1D representations, $B_1$ and $B_2$ (corresponding to two types of $f$-wave pairing). (iii) The continuum group, ${\rm O}(2)$, has an infinite set of 2D real representations, classified by odd orbital momenta, $l = 1,3,5,\ldots$.

We now consider a pairing channel corresponding to a 2D representation of $\mathbb{T} \otimes \mathbb{G}$. There are two real eigenfunctions for this representation: $\phi_{1}({\bf k})$ and $\phi_{2}({\bf k})$. If the order parameter is proportional to either of them, it is real and corresponds to a topologically trivial state with zero winding number. We prove below that such a state is always unstable. The invariance of the Hamiltonian under $\mathbb{T} \otimes \mathbb{G}$ ensures ${\cal F}_{\rm Non-top}\! = {\cal F}\left[ \phi_{1}({\bf k})\right]={\cal F} \left[\phi_{2}({\bf k})\right]$ ({\em e.~g.}, $p_x$- and $p_y$-states have the same energies in continuum).

Let us show that one can always construct a new TRSB state with
\[
\phi_{\rm  TRSB}({\bf k})=\frac{1}{\sqrt{2}}\left[ \phi_{1}({\bf k})+i \phi_{2}({\bf k})\right]
\]
that has a lower free energy than ${\cal F}_{\rm Non-top}$.  One can see from Eq.~(\ref{F2}) that ${\cal F}_{II} \left[ \phi_{\rm  TRSB}({\bf k}) \right] = {\cal F}_{II} \left[\phi_{1}({\bf k})\right] =  {\cal F}_{II} \left[\phi_{2}({\bf k})\right] $ because $\phi_{\rm  TRSB}^2({\bf k})=\phi_{1}^2({\bf k})/2+ \phi_{2}^2({\bf k})/2$,.
To handle the less trivial ``quasiparticle part'' of the free energy (\ref{F1})
 we take advantage of the Jensen's inequality which states that for any function with
 $f''(x)<0$, $f(x/2+y/2)<f(x)/2+f(y)/2$ for any $x \ne y$. The integrand in Eq.~(\ref{F1}) for ${\cal F}_{I}$
 is a concave function of $x= |\Delta_\mathbf{k}|^2$ and
 therefore satisfies the Jensen's inequality (which after integration over momentum becomes a
 strong inequality for all physically relevant cases).
 Since $\phi_{\rm  TRSB}^2({\bf k})=\phi_{1}^2({\bf k})/2+ \phi_{2}^2({\bf k})/2$, we have proven that
\begin{equation}
\label{ineq}
{\cal F} \left[\phi_{\rm  TRSB}({\bf k}) \right] < \frac{
{\cal F} \left[ \phi_{1}({\bf k}) \right] + {\cal F} \left[ \phi_{2}({\bf k}) \right] }{2}
\equiv
{\cal F}_{\rm Non-top}.
\end{equation}
This inequality (to which we refer to as ``theorem'')
represents the main result of our work and proves that a TRSB phase is always energetically favorable within a
2D representation. This is a strong statement that is completely independent of microscopic details, such as hoppings and interactions, and relies only symmetry. It leads, in particular, to the conclusion that any single-band spinless SC (and certain models of spin-polarized SCs) originating from a square lattice with singly-connected FS must be a $(p + ip)$-paired state. Similarly, any SC arising from
spinless fermions in continuum must be of a $(2l +1) + i(2l +1)$-type, which is topologically nontrivial. This includes all continuum models with attractive forces and conceivably some continuum models with weak repulsion that
may give rise to pairing via Kohn-Luttinger mechanism~\cite{Kohn1965,Chubukov1993,Galitski2003}.
Since a large number of lattice fermion Hamiltonians at low particle densities reduce to an effective single-band continuum model,
it means that at least in this low-density regime any paired state is guaranteed to be topological.



\section{Lattice Models}
\label{sec:lattice}
To illustrate how our theorem manifests itself in practice, we examine specific  models within a large class of generic tight-binding Hamiltonians on a lattice
\begin{equation}
\hat{\mathcal{H}} =-\sum_{\mathbf{r},\mathbf{r}'} t_{\mathbf{r},\mathbf{r}'} \hat{c}^\dagger_{\mathbf{r}} \hat{c}_{\mathbf{r}'}
-\mu \sum_{\mathbf{r}} \hat{c}^\dagger_{\mathbf{r}} \hat{c}_{\mathbf{r}}
+ \sum_{\langle{\mathbf r}, {\mathbf r}'\rangle} V_{{\mathbf r}, {\mathbf r}'}  \hat{c}^\dagger_{\mathbf{r}} \hat{c}^\dagger_{\mathbf{r}'}
\hat{c}_{\mathbf{r}'}\hat{c}_{\mathbf{r}},
\nonumber
\end{equation}
where $\hat{c}^\dagger_\mathbf{r}$/$\hat{c}_\mathbf{r}$ creates/annihilates a fermion on a lattice site $\mathbf{r}$. We note that this
real-space Hamiltonian reduces to a more general model (\ref{H0}) via a lattice Fourier-transform. For the sake of concreteness,
we focus below on the following two  models with nearest-neighbor hoppings, $t_{{\bf r},{\bf r}'} = t \delta_{|{\bf r} - {\bf r}'|,1}$
and nearest-neighbor attraction, $V_{{\bf r},{\bf r}'} = - g \delta_{|{\bf r} - {\bf r}'|,1}$ on (i)~a simple square lattice and (ii)~a simple triangular lattice.

\subsection{Square Lattice}
\label{sec:squ}
The square lattice case corresponds to the  $D_4$ symmetry group, which has only a 2D representation.
The attractive interaction
guarantees that the ground state is a SC~\cite{ShankarRG}  and the general theorem~(\ref{ineq}) guarantees
that it is topologically non-trivial. To see how this happens in the specific model,
we define two independent order parameters on horizontal and vertical links:
$\Delta_{n}=g\avg{\hat{c}_\mathbf{r}\hat{c}_{\mathbf{r}+{\bm e}_n}}$,
where $n=x$ or $y$ and $\boldsymbol{e}_{n}$ is the corresponding lattice vector (we use units where the lattice constant, $a=1$).
These real-space order parameters  are related to the momentum-space definition~\eqref{def_op} via $\Delta_\mathbf{k}=2i \sum\limits_{\alpha = x,y} \Delta_\alpha \phi_\alpha ({\bf k})$,
with the  BCS interaction being
$\tilde{f}_{\mathbf{k},\mathbf{k}'}=-g \sum\limits_{\alpha = x,y} \phi_\alpha ({\bf k}) \phi_\alpha ({\bf k}')$.
Here we defined two eigenfunctions of the above-mentioned 2D representation of $D_4$:
$\phi_{x,y} ({\bf k}) = \sin\left( \mathbf{k}\cdot {\bm e}_{x,y} \right)$.
\begin{figure}
\begin{center}
\includegraphics[width=0.4\textwidth]{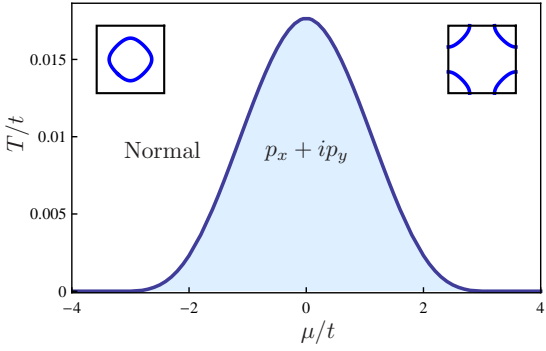}
\end{center}
\vspace*{-0.2in}
\caption{
(Color online) The phase diagram for fermions on a square lattice with nearest-neighbor hoppings and
attraction ($g/t$=1). The phase boundary separates a normal metal  and a topological $(p_x + ip_y)$-wave SC.
The insets display FSs for $\mu<0$ (left) and $\mu>0$ (right).}
\label{fig:phase_square}
\end{figure}
 It is straightforward to calculate the BCS free energy given by Eqs.~(\ref{F1}) and (\ref{F2})
 for all possible order parameters encompassed by the linear combinations
 $\Delta_\mathbf{k} = g \left[ \lambda_x \phi_x({\bf k}) +  \lambda_y \phi_y({\bf k}) \right]$,
 with arbitrary  $\lambda_{x,y} \in \mathbb{C}$. We find that a $(p+ip)$-superconducting
 state with $\lambda_x = \pm i \lambda_y$ is selected at all $\mu$.  Fig.~\ref{fig:phase_square}
 summarizes the phase diagram of the model on the $\mu - T$ plane. The maximum $T_{\rm c}$
 within the mean-field treatment occurs at half-filling.
 The tails of the particle-hole symmetric phase boundary correspond to small
 ``electron'' and ``hole'' densities, and therefore to continuum limit with the
 isotropic quadratic dispersion,  $\xi_{\bf k} = \left(k^2 - k_{\rm F}^2\right)/(2m^*)$,
 the effective mass, $m^* = 1/(2ta^2)$, and the Fermi momentum,
 $k_{\rm F} a= \sqrt{ \left| \mu \pm 4 t \right|/(2t)}$.

 It is useful to consider the continuum limit $\left| \mu \pm 4 t \right|/t \to 0$
 in more detail, as it gives a valuable insight into stability of the topological phases.
  For this purpose, we  use standard perturbative expansion~\cite{VGGL} in Eqs.~(\ref{F1}) and (\ref{F2}) to derive the Ginzburg-Landau
  free energy (per unit area):
 \begin{equation}
 \label{FGL}
\frac{1}{\cal A}{\cal F}_{\rm GL} \left[ \Delta_0, {\cal S} \right] = \nu \left( T/T_{\rm c} - 1 \right) \Delta_0^2 +  \frac{7 \zeta(3) \nu}{8 \pi^2 T^2}\, {\cal S}^{-1} \Delta_0^4,
 \end{equation}
 where $\nu = m^*/(2\pi)$ is the density of states at the FS, $T_{\rm c}$ is the BCS transition temperature, $\zeta$ is the Riemann zeta-function, $\Delta_0 = g\sqrt{ \left| \lambda_x\right|^2 + \left| \lambda_y \right|^2}$ is the modulus of the order parameter, ${\cal A}$ is the area of the sample, and  we introduced a symmetry factor, ${\cal S}$, as follows [below,  $\theta_{\bf k} = \tan^{-1}{(k_y/k_x)}$]
  \begin{equation}
 \label{S}
 {\cal S}^{-1} = \oint\limits_{{\bf k} \in {\rm FS}} \frac{d \theta_{\bf k}} {2 \pi }
\frac{\left| \lambda_x \phi_x({\bf k}) +  \lambda_y \phi_y({\bf k}) \right|^4} {\left|\lambda_x\right|^2 +
 \left|\lambda_y\right|^2}.
 \end{equation}
 The minimal free energy below $T_{\rm c}$ is given by
 ${\cal F}_{\rm  GL, min} /{\cal A} = -  {\cal S}_{\rm max}\left[ 4 \pi \nu T_{\rm c}^2 \ln^2{(T/T_{\rm c})} \right]/ [7 \zeta(3)] $. Therefore, the absolute minimum is achieved by {\em  maximizing the symmetry factor}, ${\cal S}$. In the continuum limit $|{\mathbf k}| a \to 0$, we can approximate the normalized eigenfunctions of $D_4$, by $\phi_x(\mathbf k) = \sqrt{2} \cos(\theta_{\mathbf k})$ and $\phi_y(\mathbf k) = \sqrt{2} \sin(\theta_{\mathbf k})$. Hence, the topologically trivial $p_x$- and $p_y$-states lead to ${\cal S}_{p_{x,y}} = \left\langle 4 \cos^4
{\theta_{\mathbf k}} \right\rangle_{\rm FS}^{-1} = 2/3$,
while the topological states $p_x \pm i p_y$ yield
${\cal S}_{p_{x} \pm i p_y} =\left\langle \left| e^{\pm i \theta_{\mathbf k}} \right|^4
\right\rangle_{\rm FS}^{-1} = 1>2/3$ and therefore are selected by energetics.
This fact is a special case of our general theorem summarized by Eq.~(\ref{ineq}).

We note that the mean-field BCS-type model can formally be considered for
the extreme values of the non-interacting chemical potential $|\mu|> 4t$, which is not associated with a
non-interacting FS. Hence, mean-field paired states in this limit
are not topological and correspond to the strong-pairing (Abelian) $(p + ip)$-phase
considered by Read and Green~\cite{read_prb'00}. While such a
mean-field BCS model is sensible in the context of the quantized Hall state,
it may be unphysical for fermion lattice models. Indeed,
the chemical potential, $\mu$, is renormalized by non-BCS
interactions or equivalently by superconducting fluctuations
originating from the terms with  ${\bf q} \ne {\bf 0}$ in Eq.~(\ref{H0}).
These strong renormalizations are bound to shift $\mu$
towards the physical values with a reasonable  Fermi surface,
which in a metal is guaranteed by Luttinger theorem. Hence, it is not clear whether the Abelian
$(p+ip)$ superconducting states may survive beyond mean-field. Due to these arguments,
we disregard here such case of non-topological  $(p+ip)$-paired states.

{We now derive Bogoliubov-de Gennes equations from the lattice model. These equations are often the starting point of discussions on bound states in a vortex core~\cite{Kopnin_PRB'91, Gurarie_PRB'07} and edge states~\cite{Stone_PRB'04}.} To do so, we first present the fermionic mean-field BCS Hamiltonian on a lattice as follows:
\[
\hat{\mathcal{H}}^{\mathrm{MF}}=\frac{1}{2}\sum_{\vr\vr'}\left(\hat{c}^\dag_\vr h_{\vr\vr'}\hat{c}_{\vr'}-
\hat{c}_{\vr'} h_{\vr\vr'}\hat{c}^\dag_{\vr}+\Delta_{\vr\vr'}\hat{c}_\vr\hat{c}_{\vr'}+\text{h.c.}\right)
\]
which is a real space version of Eq.~\eqref{HBCS} where $\Delta_{\vr\vr'}\equiv g\avg{\hat{c}_\mathbf{r}\hat{c}_{\mathbf{r}'}}$ is the order parameter on the bond $(\vr\vr')$ and $h_{\vr\vr'}=-t\delta_{|\vr-\vr'|,1}-\mu\delta_{\vr\vr'}$ is the matrix element of the single-particle Hamiltonian. We then follow the standard route and introduce Bogoliubov's transform $
\hat{c}_\vr=\hat{\gamma}u_\vr+\hat{\gamma}^\dag v^\ast_\vr $ and the commutation relation $[\hat{\mathcal{H}}^{\mathrm{MF}},\hat{\gamma}]=-E\hat{\gamma}$. This yields gives the desired BdG equations
\begin{equation}
\begin{split}
Eu_\vr&=\sum_{\vr'}\left(h_{\vr\vr'}u_{\vr'}+\Delta_{\vr\vr'}v_{\vr'}\right)\\
Ev_{\vr}&=\sum_{\vr'}\left(-\Delta^*_{\vr\vr'}u_{\vr'}-h_{\vr\vr'}v_{\vr'}\right)
\end{split}
\label{bdg}
\end{equation}
In principle, the order parameter $\Delta_{\vr\vr'}$ should be determined via solving BdG equation self-consistently. However, we know that in a homogeneous ground state the order parameter has a $(p+ip)$-wave pairing symmetry, i.e., $\Delta_y=\pm i\Delta_x$. If there are inhomogeneities in the system (e.g., vortices, domain walls) the pairing symmetry (associated with  the relative phase between $\Delta_y$ and $\Delta_x$ components) is not necessarily $p+ip$. But since this pairing symmetry is selected by energetics, we expect such deviation to be irrelevant for low energy physics. Therefore we can assume that the relation $\Delta_y=\pm i\Delta_x$ holds for general configurations of order parameter at the mean-field level. This is equivalent to separation of the Cooper pair wave function into parts corresponding to the center-of-mass motion and relative motion.

Now we take the continuum limit of \eqref{bdg}: $\sum_{\vr'}h_{\vr\vr'}u_{\vr'}\rightarrow \hat{\xi}(-i\nabla)u(\vr)=(-\nabla^2/2m^*-\tilde{\mu})u(\vr)$,  where $m^*$ is the effective mass and $\tilde{\mu}=\mu+4t$ is the chemical potential measured from the bottom of the band . To treat the off-diagonal part, we formally represent the second term in Eq.~(\ref{bdg}.1) as follows $\sum_{\vr'}\Delta_{\vr\vr'}v_{\vr'}=\hat{\Delta}v(\vr)$, with the gap operator being \begin{equation}
\hat{\Delta}=\sum_{\vr'}\Delta_{\vr\vr'}e^{(\vr'-\vr)\cdot\partial_{\vr}}.
\label{eq:gap_op}
\end{equation}
The order parameter $\Delta_{\vr\vr'}$, which ``lives'' on bonds, should be casted into only site-dependent form as follows:
\begin{equation}
\Delta_{\vr\vr'}=\Delta\left(\frac{\vr+\vr'}{2}\right)\exp(i \theta_{\vr'-\vr})
\end{equation}
where $\theta_{\vr'-\vr}$ is the polar angle of $\vr'-\vr$. Then, we expand \eqref{eq:gap_op} to first order in $|\vr'-\vr|=a$ and obtain the familiar BdG equations in continuum:
\begin{equation}
\begin{split}
Eu(\vr)&=\hat{\xi}(-i\nabla)u(\vr)+\hat{\Delta}v(\vr)\\
Ev(\vr)&=\hat{\Delta}^\dag u(\vr)-\hat{\xi}(-i\nabla)v(\vr)
\end{split}
\label{bdg_cont}
\end{equation}
where the gap operator $\hat{\Delta}=a\{\Delta(\vr),\partial_x+i\partial_y\}$.
An interesting question to  be addressed elsewhere is whether fluctuations and in particular deviations of pairing symmetry from $p+ip$ play a role in the topological properties.

\subsection{Triangular Lattice}
\label{Trian}

We now address the very interesting case of a simple triangular lattice. Here the $D_6$ symmetry group has both a 2D representation ($p$-wave) and two 1D representations ($f$-wave). Therefore, non-topological $f$-wave states are allowed. Low ``electron'' densities correspond to a single circular-shaped Fermi surface and must lead to the  $p + ip$-wave pairing per the same argument as above. However, the spectrum of the model is not particle-hole symmetric and at large fillings (with $\mu > \mu^*=2 t$), the electron Fermi surface splits into two hole-like Fermi pockets and maps onto an effective continuum model but with two fermion species:
\begin{equation}
\label{2species}
\hat{\cal H}_{\rm 2h, eff}= \int\limits_{\mathbf{k}} (\xi_{\mathbf{k}} \hat{h}^\dagger_{+,\mathbf{k}} \hat{h}_{+,\mathbf{k}}
+\xi_{-\mathbf{k}} \hat{h}^\dagger_{-,\mathbf{k}} \hat{h}_{-,\mathbf{k}})+\textrm{interactions},
\end{equation}
where $\hat{h}_{\pm,\mathbf{k}}$ are  fermion operators near the two pockets labeled by a pseudospin index $\sigma=\pm$ and the
spectrum is asymptotically given by
\begin{equation}
\xi_{\mathbf{k}}=k^2/2m+ \alpha(k_x^3-3 k_x k_y^2) -  E_{\rm F},
\label{xi_tri}
\end{equation}

with ${\bf k}$ measured from the
corner points of the hexagonal Brillouin zone. Note that under a $\pi$ or
$\pm \pi/3$ rotation, the spectrum  transforms as $\xi_{\mathbf{k}}\rightarrow \xi_{-\mathbf{k}} $  and this symmetry is preserved if $\sigma \rightarrow  -\sigma$. This leads to a pairing analogous  to the $s$-wave pairing of spin-$1/2$ fermions, with the order parameter of the inter-pocket pairing defined as $\Delta_h = g\int_{\bf k}\avg{\hat{h}_{+,\mathbf{k}}\hat{h}_{-,\mathbf{-k}}}$. However, this is an $f$-wave pairing state, because under a $\pi/3$-rotation, $\sigma \rightarrow -\sigma$ and $\Delta_h({\bf k})$ changes sign.
\begin{figure}
\begin{center}
\includegraphics[width=0.4\textwidth]{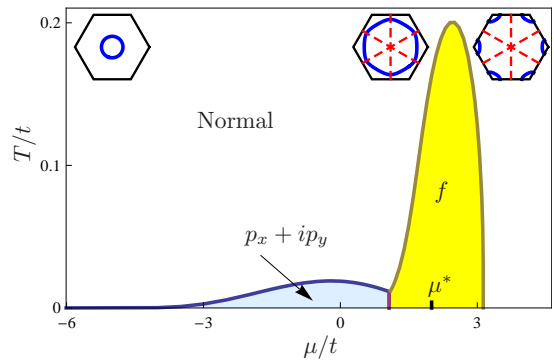}
\end{center}
\vspace*{-0.2in}
\caption{(Color online) The phase diagram for spinless fermions on a triangular lattice with nearest-neighbor hoppings  and
attraction ($g/t$=1). The bottom of the band is located at $\mu=-6t$ and the top is at $\mu=3t$; $\mu^*=2t$ corresponds to a van Hove singularity.
Two SC phases, with $(p_x+i p_y)$- and $f$-wave symmetries are present. They are separated by a first-order phase transition
at $\mu_{\rm cr}/t \approx 1.057$. The insets (left to right) are the FSs for $\mu<\mu^*$,
$\mu\lesssim\mu^*$, and $\mu>\mu^*$ and the dashed lines indicate the nodal directions of the $f$-wave SC.}
\vspace*{-0.1in}
\label{fig:phase_triangle}
\end{figure}
Since the low-density limit leads to a topological  phase and the high-density limit leads to an $f$-wave topologically trivial state, there must be a quantum phase transition in between. The entire phase diagram can be derived using Eqs.~(\ref{F1}) and (\ref{F2}) and the real-space construction as follows:
On a triangular lattice we can define three order parameters on the nearest neighbor
bonds corresponding  to the three lattice vectors, $\boldsymbol{e}_n$
with azimuth angles $2 n \pi/3$ and $n=0,1,2$:
$\Delta_{n}=g\avg{\hat{c}_\mathbf{r}\hat{c}_{\mathbf{r}+\boldsymbol{e}_{n}}}$.
Two different types of pairing channels are formed by these three order parameters:
An $f$-wave channel with $\Delta_{n}=\Delta$ and  a $p$-wave channel
with $\Delta_{n}=\Delta e^{\pm2 \pi i n/3}$.

The resulting phase diagram is shown in Fig.~\ref{fig:phase_triangle}. As expected, a topological $p+ip$-wave  SC state with $\Delta_{n}=\Delta e^{\pm 2 \pi i n /3}$ is stabilized at low fillings, while an $f$-wave state with $\Delta_{n}=\Delta$ is favored at high densities. These phases are separated by a first-order transition. As shown in Fig. \ref{fig:phase_triangle}, the van Hove singularity $\mu=\mu^*$ gives rise to a maximal $T_{\rm c}$ and is located inside the $f$-wave superconducting dome. This point represents another type of a quantum  transition that separates two qualitatively different topologically trivial paired states: (1)~For $\mu<\mu^*$, there is just one electron-type Fermi pocket  that is cut by the nodes of the  $f$-wave gap in the directions, $\theta_{\rm node}^{(m)} = m \pi/3+\pi/6$. This gives rise to gapless quasiparticles. (2)~For $\mu>\mu^*$, no FS can be cut and the nodal quasiparticles disappear. The phase becomes  fully gapped and eventually crosses over to the two-specie continuum model~(\ref{2species}). Experimentally, the two types of $f$-wave phases can be distinguished  by  different $T$-dependence of the heat capacity.

{We also present the Bogoliubov-de Gennes equations for the $f$-wave pairing state in high-density limit  $\mu\rightarrow 3t$.} Their derivation goes along the same lines as that given in Section~\ref{sec:squ} for $p_x+ip_y$ pairing SC. However, in the $f$-wave  case, the momentum space order parameter is given by $\Delta_\vk=\Delta(\sin\vk\cdot \vec{e}_1+\sin\vk\cdot \vec{e}_2+\sin\vk\cdot \vec{e}_3)$. Therefore, the order parameter reads:
\[
\Delta_{\vr\vr'}=\Delta\left(\frac{\vr+\vr'}{2}\right)\cos(3 \theta_{\vr'-\vr}).
\]
Since $\vec{e}_1+\vec{e}_2+\vec{e}_3=0$, the leading term in the expansion is $\sim a^3$.
With some algebra one can show that the gap operator is
\begin{equation}
\hat{\Delta}=\frac{a^3}{24}\sum_{n=0}^2\{\partial_n,\{\partial_n,\{\partial_n,\Delta(\vr)\}\}\}
\end{equation}
with $\partial_n\equiv\nabla\cdot \vec{e}_n$ and the BdG equation takes the form of Eq.~\eqref{bdg_cont}.

\section{Discussion and Conclusions}
{
In conclusion, we discover that topological superconducting phases breaking time -reversal symmetry emerge naturally within a large class of spinless fermion models.
The technique we apply here has a close relation to BCS mean field theory of a spin-triplet superfluid $^3$He~\cite{Leggett_RevModPhys'75,Vollhardt_book'1990}, which concluded that the B-phase with
isotropic gap is stabilized compared to anisotropic A-phase~\cite{helium}. However, we have shown that
similar conclusion can be generalized to any band structures, filling factors, and interactions, as long as
the system satisfies proper (discrete) rotational group symmetries. More importantly, our proof is insensitive
to the existence of the ``nodes''. In continuum, it has been argued that a $p_x$ state is unstable against the
$p_x+i p_y$ pairing state, because the former has nodes thus having smaller condensation energy. However,
the stability of a nodeless $p_x$ state, which could exists in lattice models, was unclear before this our work.}

{
We should also emphasize that although the discussions above focus on spinless fermions, all the conclusions
can be generalized to the triplet pairing channels of spin-$1/2$ fermions, because these pairing channels
also correspond to the space-inversion odd representations of the symmetry group. In addition, we note
that any pairing state that spontaneously breaks a lattice rotational symmetry must have at least one
degenerate state for both spinless and spin-$1/2$ fermions. Our theorem
indicates that these type of states must have a complex pairing order parameter to be energetically stable.
}

This work is supported by DARPA-QuEST, JQI-NSF-PFC, and US-ARO.


\begin{thebibliography}{24}
\bibitem[{hel()}]{helium}
\bibinfo{note}{The A-phase can be stablized under high pressure. In this case,
  due to strong spin-fluctuations in liquid $^3$He, the ground state energy
  departs from the BCS theory, which is not a contradiction to our conclusion.}
\end{thebibliography}

\end{document}